\documentclass[11pt,a4paper]{article}

\usepackage[hyperref]{emnlp2020}
\usepackage{times}
\usepackage{latexsym}

\aclfinalcopy % Uncomment this line for the final submission
\usepackage{subcaption}
\usepackage{graphicx}
\usepackage{mathtools}  
\usepackage{booktabs}
\usepackage{xspace}
\usepackage{xcolor}
\usepackage{comment}
\usepackage{soul}
\usepackage{url}

\usepackage{enumitem}
\usepackage{arydshln}
\usepackage{fdsymbol}

\newcommand{\latinword}[1]{\texttt{#1}}%

\newcommand{\model}{\latinword{SciSight}\xspace}

\author{
Tom Hope$^{\clubsuit,\heartsuit}$
~~~~~~~~~Jason Portenoy$^{\clubsuit,\varheartsuit}$\Thanks{\enspace Equal contribution.}
~~~~~~~~~Kishore Vasan$^{\varheartsuit}$\textcolor{darkblue}{\footnotemark[1]}
~~~~~~~~~Jonathan Borchardt$^{\clubsuit}$\textcolor{darkblue}{\footnotemark[1]}
\\\AND
~~~~~~~~~Eric Horvitz$^{\spadesuit}$
~~~~~~~~~Daniel S. Weld$^{\clubsuit,\heartsuit}$
~~~~~~~~~Marti A. Hearst$^{\diamondsuit}$
~~~~~~~~~Jevin West$^{\varheartsuit}$
\\
$^{\clubsuit}$Allen Institute for Artificial Intelligence\\
$^{\heartsuit}$Paul G. Allen School for Computer Science \& Engineering, University of Washington\\
$^{\varheartsuit}$Information School, University of Washington\\
$^{\spadesuit}$Microsoft Research  ~
$^{\diamondsuit}$University of California, Berkeley\\
{\tt \{tomh,jasonp,jonathanb\}@allenai.org} 
~~~ ~~~{\tt \{kishorev,jevinw\}@uw.edu} \\ \\ \\
}

\title{\model: Combining faceted navigation and research group detection for COVID-19 exploratory scientific search}

\date{}

\begin{document}
\maketitle
\begin{abstract}
The COVID-19 pandemic has sparked unprecedented mobilization of scientists, generating a deluge of papers that makes it hard for researchers to keep track and explore new directions. Search engines are designed for targeted queries, not for discovery of connections across a corpus. In this paper, we present \textbf{SciSight, a system for exploratory search} of COVID-19 research integrating two key capabilities: first, exploring associations between biomedical facets automatically extracted from papers (e.g., genes, drugs, diseases, patient outcomes); second, combining textual and network information to search and visualize \emph{groups} of researchers and their ties. SciSight\footnote{\url{http://scisight.apps.allenai.org/}} has so far served over $13K$ users with over $37K$ page views and $13\%$ returns.

%   We extract entities using a language model pre-trained on several biomedical information extraction tasks, and enrich them with data from the Microsoft Academic Graph (MAG). To find research groups automatically, we use hierarchical clustering with overlap to allow authors, as they do, to belong to multiple groups. Finally, we introduce a novel presentation of these groups based on both topical and social affinities, allowing users to drill down from groups to papers to associations between entities, and update query suggestions on the fly with the goal of facilitating exploratory navigation.
\end{abstract}

%% The code below is generated by the tool at http://dl.acm.org/ccs.cfm.
%% Please copy and paste the code instead of the example below.
%%
% \begin{CCSXML}
% <ccs2012>
%  <concept>
%   <concept_id>10010520.10010553.10010562</concept_id>
%   <concept_desc>Computer systems organization~Embedded systems</concept_desc>
%   <concept_significance>500</concept_significance>
%  </concept>
%  <concept>
%   <concept_id>10010520.10010575.10010755</concept_id>
%   <concept_desc>Computer systems organization~Redundancy</concept_desc>
%   <concept_significance>300</concept_significance>
%  </concept>
%  <concept>
%   <concept_id>10010520.10010553.10010554</concept_id>
%   <concept_desc>Computer systems organization~Robotics</concept_desc>
%   <concept_significance>100</concept_significance>
%  </concept>
%  <concept>
%   <concept_id>10003033.10003083.10003095</concept_id>
%   <concept_desc>Networks~Network reliability</concept_desc>
%   <concept_significance>100</concept_significance>
%  </concept>
% </ccs2012>
% \end{CCSXML}

% \ccsdesc[500]{Computer systems organization~Embedded systems}
% \ccsdesc[300]{Computer systems organization~Redundancy}
% \ccsdesc{Computer systems organization~Robotics}
% \ccsdesc[100]{Networks~Network reliability}

%%
%% Keywords. The author(s) should pick words that accurately describe
%% the work being presented. Separate the keywords with commas.

%%
%% This command processes the author and affiliation and title
%% information and builds the first part of the formatted document.

\section{Introduction}
Scientists worldwide are racing against the growing number of COVID-19 infections, to understand and treat the disease \cite{nytimes}. However, a very different kind of exponential growth has been plaguing researchers -- the flurry of papers published every year, at a rate that continues to increase \cite{williamson2019exploring}. At the time of this writing, the COVID-19 Open Research Dataset (CORD-19) \cite{wang2020cord} includes over 130,000 publications of potential relevance, both historical and cutting-edge.

To boost scientific discovery over this corpus, we propose SciSight, a working prototype system for \textbf{exploratory search of the COVID-19 literature}. Unlike many tools (see Section \ref{sec:related}), we shift the focus from searching over lists of papers or authors, to navigating \textbf{networks of biomedical concepts and research groups} -- for example, exploring links between COVID-19 and other diseases, or labs working on treatments. While search engines are a powerful tool for finding documents, they are mostly geared toward targeted search, when researchers know what they are looking for -- less useful for exploring connections that are not obvious from reading individual papers \cite{bales2009evaluation,white2009exploratory}.

Building exploratory interfaces in science is difficult not only due to the complexities of scientific content, but also because of social undercurrents that have tremendous effects on the construction of knowledge \cite{wagner2005network, pan2012world, west2013role} (as reflected, for example, in biased citation patterns \cite{king2017men}). Silos of knowledge throughout the literature \cite{vilhena2014finding}\footnote{So Long to the Silos, Nature Biotech, \url{https://www.nature.com/articles/nbt.3544}} can hinder research advancement and cross-fertilization across groups and fields that is crucial for driving innovation \cite{hope2017accelerating,kittur2019scaling}, ultimately impacting human lives \cite{loevinsohn2015cost}. These problems are acute when it comes to the COVID-19 pandemic, with new information rapidly emerging and urgently needed.

We aim to incorporate the social structure into an intuitive design interface, to help researchers \textbf{make connections to other groups and ideas in the literature} by traversing across networks of concepts and groups of scientists -- helping users discover \textbf{who is working on what, and where?}

We identify groups by clustering the co-authorship graph, and extract topics and entities from the group's papers. Each group is represented using textual and network information: the group's salient  authors (\emph{who}), the topics they work on (\emph{what}) and their affiliations (\emph{where}). We build meta-edges capturing topical affinity between groups using a language model fine-tuned for semantic similarity, and present approaches for searching for groups with queries consisting of authors, topics and affiliations. Each selected query automatically suggests new queries to try, to support exploration \cite{kairam2015refinery}.

In summary, our main contributions:
\begin{itemize}
  \item A working prototype for exploratory search and visualization of COVID-19 scientific literature and collaboration networks, based on a fusion of automatically extracted textual information (topics, entities) and co-authorship network information. 
  
  \item  User interviews with experts suggest that SciSight can help complement standard search and help discover new directions.
  
\end{itemize}

\section{Related work}
\label{sec:related}

The field of bibliometric visualization goes back decades \cite{borgman2002scholarly}, with a large body of work. Visualizations of the scientific literature can take many shapes and forms, with the aim of depicting the connections between fields, topics, authors, and, most commonly, papers \cite{balesbibliometric}. While much research has been done in this field over the years, actual tools that are readily available primarily focus on visualization of citation-based graphs between \emph{individual} authors, papers or topics \cite{van2010software,synnestvedt2005citespace,persson2009use}. While this rich information could in theory be useful, in practice it often renders the visualization inscrutable, especially for real-world networks comprising many authors. This problem is especially acute when the goal is to enable discovery of new areas with unfamiliar authors. 

Recently, such tools have been applied to COVID-19 papers, such as journal networks and heat maps of frequently occurring terms \cite{haghani2020scientific}. However, many tools require training before being able to be used, and state of the art bibliometric mapping is currently considered ``complex and unwieldy'' \cite{balesbibliometric}, potentially because the typical user ``does not \textit{immediately} comprehend a map and (as a result) is not enticed into using it'' \cite{buter2006combining}.

% As a representative example, we briefly review VOSviewer \cite{van2010software}, a popular bibliometric visualization tool supporting network visualizations including author clustering. While VOSviewer does not offer an interface dedicated to COVID-19, we download COVID-19 paper metadata from Elsevier's Scopus \cite{elsevier}, and then upload to the VOSviewer co-authorship graph visualizer in its supported data format. Figure \ref{fig:vos} shows the results. Nodes represent authors, edges represent a co-authorship relation, and colors denote author clusters obtained with community detection (in this case resulting in 13 clusters). While paper metadata was loaded into the tool, the tool did not support searching for specific keywords or author affiliations, or viewing relevant papers and metadata. We observed similar features in other free and commercial tools \cite{synnestvedt2005citespace,persson2009use, wong2004spire}.

\textbf{COVID-19 tools}. In response to COVID-19, many tools for exploring the relevant literature have been released. The great majority featured paper search interfaces, with lists of titles and abstracts being the main focus. Many of the COVID-19 tools we reviewed included standard faceted search functionality~\cite{yee2003faceted,hearst2006faceted,tunkelang2009faceted}, enabling users to filter papers according to various facets. In a search tool from Microsoft Azure \cite{Microsoft}, for example, users can filter search results by various facets (such as by authors or gene mentions extracted automatically from texts). Similar services were made available by IBM Watson \cite{IBM}, Elsevier \cite{elsevier} and the National Institutes of Health \cite{litcov}.

Currently a small number of tools focus on concept associations. One tool \cite{semviz} feeds a COVID-19 knowledge graph (KG) from \cite{blender} into Kibana\footnote{https://www.elastic.co/kibana}, an external product for creating dashboards with complex heat maps of term frequencies in documents, including a specialized query language for users with sufficient familiarity with Kibana. A tool from \cite{bras2020visualising} shows clusters of high-level topics extracted with Latent Dirichlet Allocation (LDA) \cite{blei2003latent} (visualized with word clouds).  

In this paper, we integrate textual information from papers and the network of author collaborations, allowing users to drill down from research groups to papers to associations between entities in one system, with a custom interface aimed to help users ``comprehend the map'' \cite{balesbibliometric,buter2006combining} intuitively.

% \begin{figure*}
% \centering
% \includegraphics[width=.48\textwidth]{figs/mag1.png}
% \includegraphics[width=.48\textwidth]{figs/ibm.png}

% \medskip
% \includegraphics[width=.32\textwidth]{figs/elsv.png}
% \includegraphics[width=.32\textwidth]{figs/litcovid.png}
% \includegraphics[width=.32\textwidth]{figs/cvodischol.png}
% \caption{COVID-19 faceted search interfaces, focusing on papers. From upper left, clockwise: Screenshots of search tools from Microsoft\cite{Microsoft}, IBM\cite{IBM}, Elsevier\cite{elsevier}, NIH\cite{litcov}, Berkeley\cite{covidscho}.}
% \label{fig:relatedsearch}
% \end{figure*}

% \begin{figure*}
%   \includegraphics[width=\linewidth]{figs/covid searches.png}
%   \caption{COVID-19 faceted search interfaces, focusing on papers. From upper left, clockwise: Screenshots of search tools from Microsoft\cite{Microsoft}, IBM\cite{IBM}, Elsevier\cite{elsevier}, NIH\cite{litcov}, Berkeley\cite{covidscho}. \tnote{add higher res}}
%   ~\label{fig:relatedsearch}
% \end{figure*}

% \begin{figure*}
% \centering
% \includegraphics[width=.32\textwidth]{figs/wellai.png}
% \includegraphics[width=.33\textwidth]{figs/kibana.png}
% \includegraphics[width=.32\textwidth]{figs/mag2.png}
% \caption{COVID-19 entity interaction tools. Screenshots from WellAI \cite{Wellai} (left), Brandeis \cite{semviz} (center) and Microsoft Azure \cite{Microsoft} (right).}
% \label{fig:relatedconcept}
% \end{figure*}

\section{SciSight: system overview}
\label{sec:scisight}
In this section we present an overview of our prototype and its  distinct components. We motivate each by discussing researcher needs. We illustrate SciSight's features and potential with the following illustrative example: \\

\textit{Marc is a researcher interested in exploring Chloroquine, an anti-malarial drug that has been surrounded with controversies in the context of COVID-19 \cite{touret2020chloroquine}. In particular, Marc wants to find connections between Chloroquine and other drugs and diseases, and to understand how these entities are interconnected in order to explore other candidate drugs and potential side-effects. Marc is familiar with the field and its main papers, but the amount of related work is overwhelming with a litany of drugs and diseases. Complicating things further, knowing that Chloroquine is not a new medication, Marc wants to examine connections across years of research, not just recent work.}

\subsection{Collocation explorer} 
\begin{figure}
\begin{subfigure}{0.99\linewidth}
\includegraphics[width=0.99\linewidth]{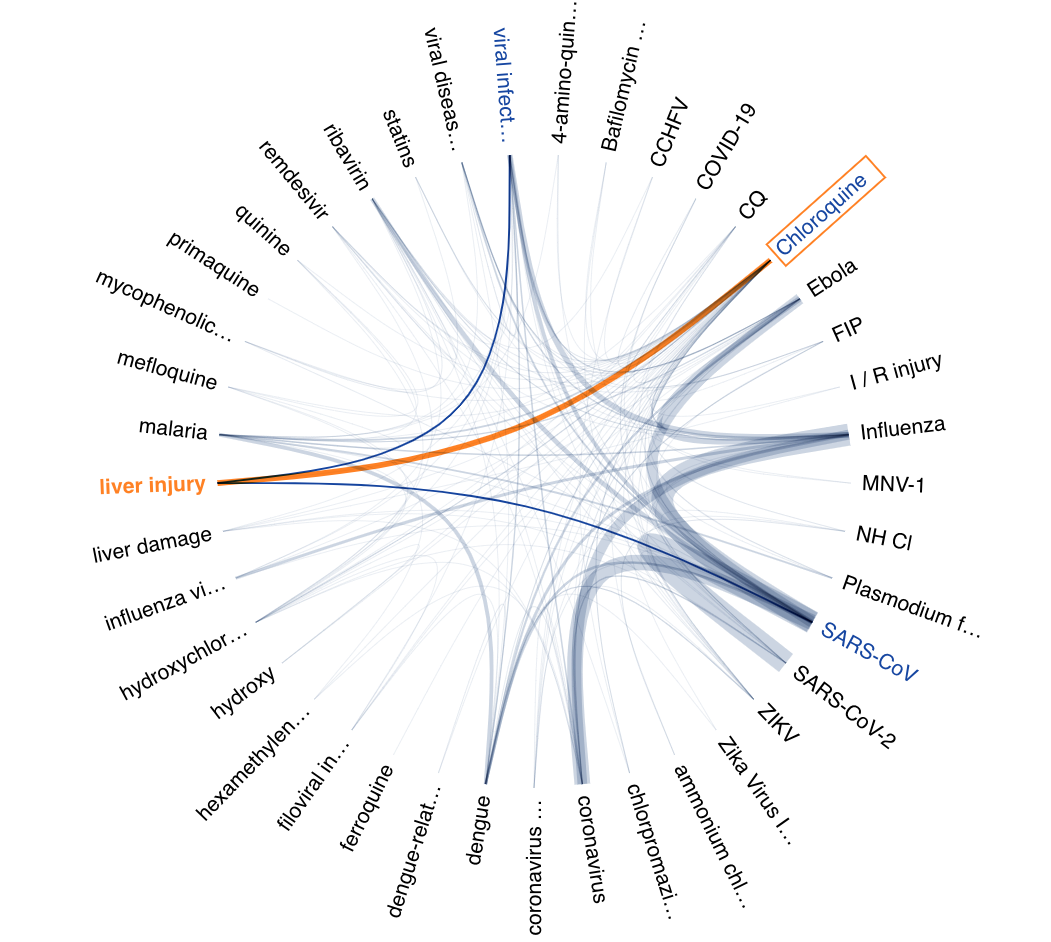} 
\label{fig:collo}
\caption{}
\end{subfigure}
\begin{subfigure}{0.95\linewidth}
\includegraphics[width=0.95\linewidth]{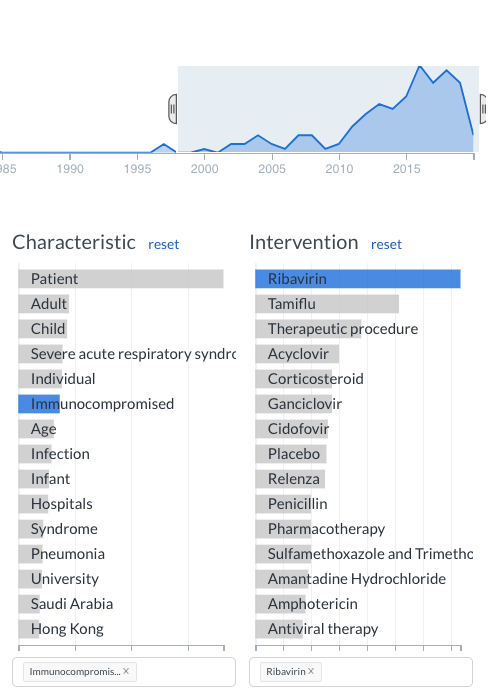}
\label{fig:expsearch}
\caption{}
\end{subfigure}
% \begin{subfigure}{0.95\linewidth}
% \includegraphics[width=0.95\linewidth]{figs/Riba.png} 
% \label{fig:facetpapers}
% \caption{}
% \end{subfigure}
\caption{(a) \textbf{Collocation explorer}: corpus-wide associations between biomedical entities, such as drugs and conditions. Highlighted in the figure is the edge between Chloroquine and liver injury. (b) \textbf{Exploratory search} of connections between patient characteristics and interventions. Papers working with immunocomprimised patients and Ribavirin would be listed below the facet feature. The time graph above shows the number of papers per year with these criteria.} %(c) \textbf{Drilling down}: All our features support drilling down to relevant papers, after the user explores different facets and converges on a query of interest.}
\label{fig:facets}
\end{figure}

Users of SciSight can search for a term/concept of interest, or get suggestions based on important COVID-19 topics. Searching for a term displays a network of top related terms mined from the corpus, based on term collocation counts across the corpus (co-appearance in the same sentence). Entities are displayed in a customized chord diagram \cite{lee2015people} layout\footnote{SciSight is implemented with React, server-side Crossfilter, DC.js, D3.js, and Varnish.}, with edge width corresponding to collocation frequency.  As seen in Figure \ref{fig:facets}a, interrelations between all terms are shown (not just with the query), presenting the user with more potential connections to explore (users can also  control the number of entities shown).  Clicking an edge between two entities displays a list of papers containing both terms.

\ 

\textit{Continuing our example, Marc can search for \textit{Chloroquine} and see its network of associations, such as a potential connection to liver damage, or its connection to other drugs such as the anti-viral drug Ribavirin. Marc can navigate the graph by clicking nodes to further explore new associations (e.g., clicking \textit{liver damage} to potentially discover more related drugs and diseases). Navigation is known to help facilitate exploration \cite{kairam2015refinery}, such as when users do not have a pinpointed query in mind \cite{white2009exploratory}.} 

\

\textbf{Entity extraction and selection} To extract entities we use S2ORC-BERT \cite{lo-wang-2020-s2orc}, a new language model pre-trained on a large corpus of scientific papers. This model is fine-tuned\footnote{ \url{https://github.com/allenai/scibert/blob/master/scripts/exp.py}.} on two separate biomedical named entity recognition (NER) tasks (BC5CDR \cite{li2016biocreative} and JNLPBA \cite{kim2004introduction}), enabling us to extract spans of text corresponding to \textit{proteins, genes, cells, drugs, and diseases} from across the corpus. We extract entities only from titles and abstracts of papers to reduce noise and focus on the more salient entities in each paper. We show only entities collocated at least twice with other entities. Our choice of entities is the result of an initial round of interviews with biomedical experts, identifying these concepts as fundamental to the study of the virus. Participants with a more clinical orientation expressed interest in viewing associations between drugs and diseases, while users from a biology background wished to focus on proteins, genes and cells. When asked whether they would prefer to have all types of entities in one view, participants responded with a preference for separate graphs to avoid clutter and reduce cognitive effort. 

\subsection{Faceted exploratory search} 

Similarly to other tools, we incorporate a faceted search tool into SciSight. Our focus is on exploration of topics and associations, with relevant papers displayed below the facets for users wishing to dig deeper after refining their search -- rather than being featured front and center. When searching for a topic or an author, new suggestions to help refine the search are presented based on top co-mentions with the initial query to help prevent fixation on an initial topic and boost associative exploration \cite{kairam2015refinery}. In our prototype for this feature we aimed at providing one compact set of facets that can cater to a wide range of interests but still be sufficiently granular. Based on formative interviews and a review of biomedical concept taxonomies, we converged on three widely-used topical facets in biomedicine, that capture characteristics of patients or the problem, interventions, and outcomes \cite{schardt2007utilization} (see Figure \ref{fig:facets}b), extracted automatically from biomedical abstracts with the distant supervision model in \cite{wallace2016extracting}. In addition, CORD-19 metadata facets are available, such as journal, affiliation and author. The number of relevant publications is shown over time, possibly revealing trends for specific facets. Users can adjust the time range to update the papers and facets displayed. 

\

\textit{Having spotted a potential connection to Ribavirin, Marc searches for it under the intervention facet to find out about related patient populations and outcomes, and to see how often it has been mentioned over time (see Figure \ref{fig:facets}b). A characteristic that pops-up and catches Marc's attention is immunocompromised patients, as he recalls a colleague discussing the risk of treating such populations. He finds peaks of interest around some points in time, and drilling down to papers from around 2016 finds a paper with the following conclusion: "No consensus was found regarding the use of oral versus inhaled RBV... such heterogeneity demonstrates the need for further studies ... in immunocompromised hosts." Marc realizes his knowledge of this domain is lacking, and decides to zoom out and find out what groups and labs are working on immunity and viral diseases, perhaps also discovering some familiar collaborators.}

\subsection{Network of science} 

In the course of formative interviews with domain experts, participants expressed the need to see what other groups are doing in order to keep track, explore new fields and potentially collaborate. We build a visualization of groups and their ties and integrate this social graph with exploratory faceted search over topics, authors and affiliations. We design our tool with the following components.

\subsubsection{Author groups} To identify groups of researchers, we start by constructing a co-authorship network in which links between authors represent collaboration on a paper, weighted by the number of papers. We then employ an overlapping community detection algorithm based on ego-splitting \cite{epasto2017ego} so that authors can belong to multiple clusters (groups). We relax the assumption typically made in co-authorship analysis that authors belong to one group alone -- in reality, researchers can ``wear many hats'' and belong to different groups depending on what they work on and with whom.

\begin{figure}
\includegraphics[width=\linewidth]{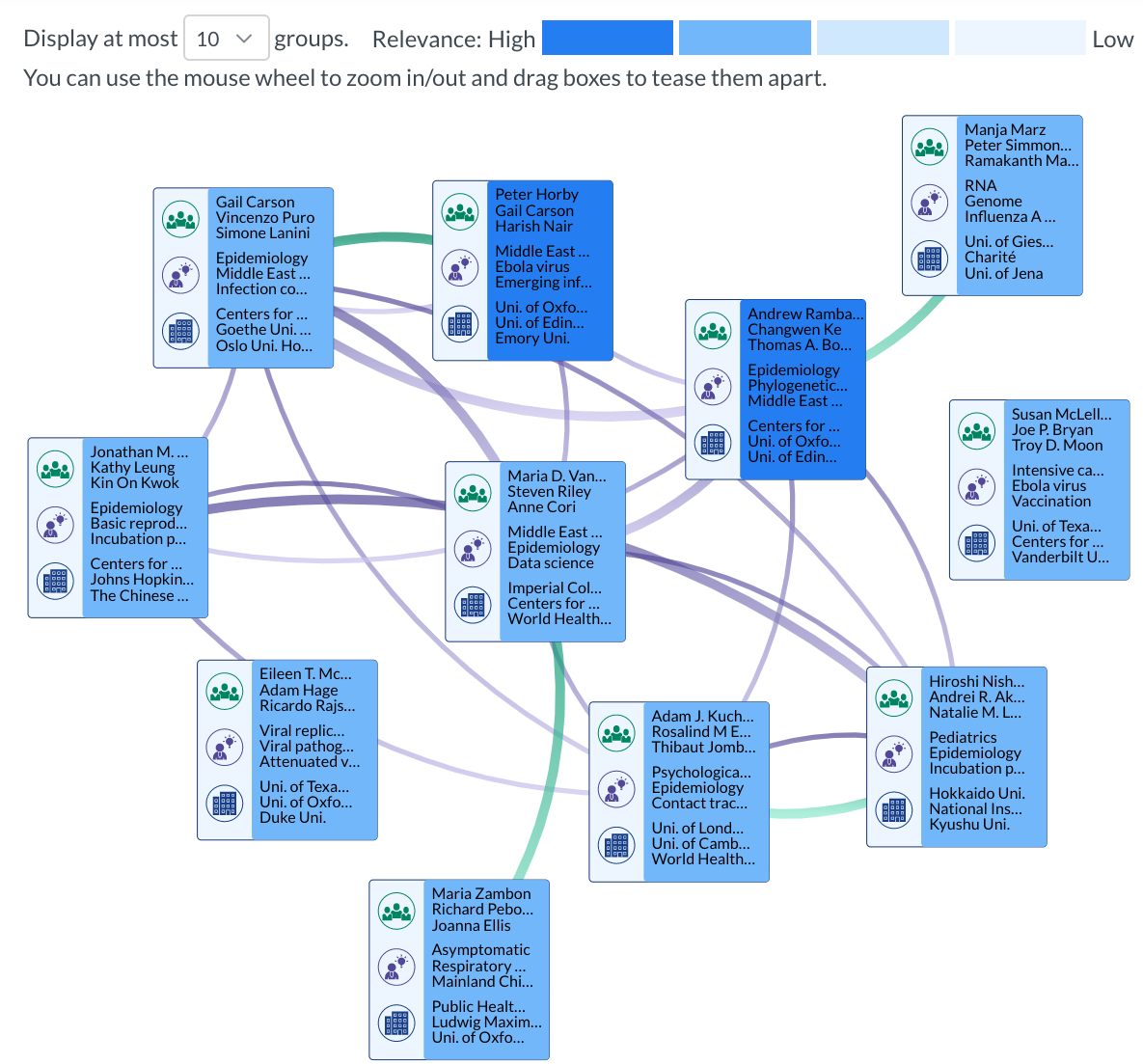}

% \begin{subfigure}{0.95\linewidth}
% \includegraphics[width=0.95\linewidth]{figs/Riba.png} 
% \label{fig:facetpapers}
% \caption{}
% \end{subfigure}

% \begin{subfigure}{0.4\textwidth}
% \includegraphics[width=0.5\textwidth]{figs/num_author_clusters.eps} 
% \caption{Cluster distribution for authors}
% \label{fig:cluster_au_stats}
% \end{subfigure}

% \begin{subfigure}{0.5\textwidth}
% \includegraphics[width=0.8\textwidth]{figs/cluster_emb_vis.eps}
% \caption{Embedding Clusters \tnote{drop ids, add 3-4 large cluster labels or legend}}
% \label{fig:clus_emb}
% \end{subfigure}
\caption\textbf{{Visualizing the network of groups} with group ``cards''. Each card has three icons denoting the top three authors, topics and affiliations, respectively. Card color indicates relevance to the search query. Green edges capture social affinity (shared authors), and purple edges capture topical affinity.}
\label{fig:net}
\end{figure}

As shown in Figure~\ref{fig:net}, we represent groups with ``cards'' \cite{bota2016playing} of salient authors, affiliations and topics (with information from Microsoft Academic Graph (MAG) ~\cite{sinha2015overview}). Cards are color-coded to reflect relevance to the user's initial query -- aiming to strike a balance between the relevance and diversity of the results shown. Users may select how many groups to view, zoom in/out, click a group to see a list of its topics, authors and papers.

% We observe that the co-authorship network consists of one giant connected component, and many much smaller separate components (see Table \ref{tab:co_author_stats}). We focus our study here only on the giant component; smaller connected components largely represented author disambiguation errors in the data, or authors who do not have enough of a presence in the literature to have any connections.

% \begin{table}
% \caption{Co-Authorship Network Statistics. We observe a large network with sparse connections. The giant component has a lower transitivity score but a higher density score than the larger network. <k>, <L> and CC denote average degree, average path length and clustering coefficient, respectively. Density denotes the ratio of existing edges and the number of possible edges in a complete graph; transitivity denotes the ratio existing triangles and possible triads (vertices with two edges).}
% \label{tab:co_author_stats}
% \begin{tabular}{l@{\hskip-1pt}ccccccc}
% \toprule
% & Num Nodes & Num Edges & <k> & <L> & CC & Transitivity & Density \\
% \midrule
% Complete Network & 86647 & 521704 & 12.04 & - & 0.89 & 0.65 & 0.0001 \\
% Giant Component & 38040 & 335711 & 17.65 & 6.35 & 0.89 & 0.598 & 0.0004 \\
% \bottomrule
% \end{tabular}
% \end{table}

To explore groups recently active in this space we select authors with at least one paper in CORD-19 since the year $2017$.  We focus on the giant connected component of this network (111,236 author nodes, 951,072 edges; smaller components typically represented disambiguation errors), and run the community detection algorithm. We observe a small number of ``super clusters'', large communities with hundreds of authors not densely linked, a well-known characteristic of community structure in real-world networks \cite{leskovec2009community}. We thus apply the clustering algorithm again within clusters with more than 120 authors to break them down further into denser groups. This results in 6,475 clusters. There are 5276 authors belonging to two groups; 6657 are in more than one cluster, and 1381 in more than two clusters.

% \begin{figure}{r}{0.95\linewidth}
% \includegraphics[width=0.94\linewidth]{figs/clust_dist_author_affil_pdf.pdf} 
% \caption{Two overlaid plots showing the number of authors (green) and number of affiliations (magenta) in clusters, respectively. We observe that a few groups have a large diversity of affiliations.}
% \label{fig:cluster_stats}
% \end{figure}
%To represent each cluster's ``card'', we aim for an intuitive and clear display of relevant information, so that users can immediately understand what they are shown with relative ease, unlike in many bibliometric visualizations \cite{balesbibliometric,bales2009evaluation}.
We display a mix of textual and social information: the most salient authors (\textit{who}), affiliations (\textit{where}) and topics (\textit{what}). We rank topics by their TF-IDF scores within a cluster, and authors and affiliations by relative frequency in a group. Users can also dig deeper into groups with two further levels of resolution. First, when hovering over a group with the cursor, users are shown a tooltip box with the top $5$ authors, affiliations and topics, with full names shown. Secondly, upon clicking a group we show full ranked lists of these entities, in addition to the group's papers ranked by recency (with title, abstract, journal and authors, including a hyperlink to read the full paper).

\subsubsection{Group links} We construct two types of links between groups. The first type (shown as purple edges) represents topical affinity across groups -- the interests they have in common based on publishing on similar topics. 
The second type of link (shown as green edges) captures social affinity between groups, meaning groups with many shared author relationships. 
By providing both kinds of links, the tool implicitly suggests potential collaborations or connections, particularly when a social connection does not currently exist alongside a topical one.
%Hovering over green links shows the top shared authors between two groups, and over purple links shows the top shared topics. 

% \textit{For example, in Figure \ref{fig:netlink} we can see three clusters Marc found after navigating around groups of interest. To find out which researchers work in both groups or what are the major topics of interest they share, Marc hovers over a purple edge to find out that two groups are both focusing on the Middle East Respiratory Syndrome (MERS), a previous coronavirus.}

\textbf{Cluster Relationships} To find topical affinity between clusters we embed topic surface forms with a language model trained to capture semantic similarity\footnote{RoBERTa-large-STS-SNLI\cite{liu2019roberta} \url{github.com/UKPLab/sentence-transformers}}. With each topic represented with its embedding, we get a vector representation of groups with a simple TF-IDF weighted average of embeddings, and compute cosine similarity.

% \begin{figure}
%   \begin{center}
%     \includegraphics[width=0.95\linewidth]{figs/covid_network_link.png}
%   \end{center}
% \caption{Link interpretability}
% \label{fig:netlink}
% \end{figure}

% We also experiment with a different method, using relatedness scores provided by Microsoft Academic. These scores are found using heterogeneous network embedding models to extract vector representations for nodes (such as topics), allowing to compute relatedness scores. We standardize the scores to be between 0 and 1 and use a TF-IDF weighted average of all pairwise distances between the top 10 cluster topics. Empirically we find in our initial experiments that both approaches perform similarly in terms of finding similar clusters. We leave to future work more rigorous comparisons. 

%\textbf{Author Embedding} We create embeddings of the global co-authorship network of 38040 authors using Node2Vec, an established method that finds representation of nodes using a Word2Vec approach. \knote{Cite node2vec} We then use this representation to find the embedding of the clusters as an average of the authors present in the cluster. This gives us a look at the cluster relationships from the lens of social affinity. 

\textbf{Finding Bridges} As a test case for demonstrating our framework's ability to find gaps and similarities across groups of researchers, we identify ``bridges'' between groups, potentially signifying structural holes \cite{burt2004structural} in the author network. We examine groups that work on \textit{data science} (MAG topic), a highly interdisciplinary field connecting researchers from multiple domains. We discover \textit{Derek AT Cummings}, a prominent biologist and epidemiologist with appointments at two different universities. We find him to be a sole shared author between two different clusters: one focusing on areas tied with virology and medical microbiology, while the other more associated with computational epidemiology. The former group has 15 authors, and the latter has 35. 

\textbf{Similarity Evaluation} In a preliminary experiment, we selected 30 random clusters and computed topical affinities to other clusters. For each group we randomly sample one cluster out of the top 3 closely related clusters, and another cluster from the bottom 50\% of farthest clusters (for network construction as shown to users, we only create links between top-most similar groups). We randomize the results and give them to a biomedical data analyst for annotation. We find that overall, we are able to correctly find pairs of research groups that work in similar areas with a 80\% precision. In future work we plan to collect validation data enabling to measure both precision and recall.

% \textbf{Meta-Edge Clusters} We create meta-edges between clusters based on topics and authors to find structural bridges. To create the author meta-edge network, we connect two clusters if they have a common top author and we create the topic meta-edge network based on top 5 common topics. The edge weights are based on the number of common topics between the clusters. We present the weighted degree distribution of these two networks in \ref{fig:meta_edge_dist} and some statistics in \ref{tab:meta_edge_stats}.

% \begin{table}[!h]
%   \caption{Meta-Edge Cluster Graph Statistics}
%   \label{tab: meta_edge_stats} %@{\hskip-25pt}
%   \begin{tabular}{l@{\hskip-11pt}ccccccc}
%     \toprule
%     & Num Nodes & Num Edges & <k> & Num  CC & Clustering & Transitivity & Density \\
%     \midrule
%     Author Meta-Edge & 2048 & 5696 & 5.56 & 13 & 0.49 & 0.11 & 0.002 \\
%     Topic Meta-Edge & 1999 & 71792 & 71.8 & 5 & 0.54 & 0.55 & 0.035 & \\
%     \bottomrule \\
% \end{tabular}
% \end{table}

% \begin{wrapfigure}{l}{0.4\textwidth}
% \includegraphics[width=0.4\textwidth]{figs/cluster_metaedge_degree.eps}
% \caption{Meta Edge Degree Distribution}
% \label{fig:meta_edge_dist}
% \end{wrapfigure}

% \tnote{Add table with results}

\subsubsection{Exploratory search for groups} 
Users can search topics, affiliations, or authors. We rank topics based on global TF-IDF scores. As in standard faceted search, queries across facets are conjunctive (e.g., gene sequencing AND Harvard), and queries within facets are disjunctive (e.g., gene sequencing OR bioassays). Each query consists of one or more choices for each facets. A selection automatically suggests new facets that are frequently associated with the original query, suggesting more groups and topics to explore.

The problem of finding relevant communities to a query has been explored to a certain extent under the rubric of \textit{community search} \cite{sozio2010community, fang2020effective}, in which given a graph $\mathcal{G}$ and a set of query nodes in the graph, the objective is to find a subgraph of $\mathcal{G}$ that contains the query nodes and is also densely connected. The problem of community search in \emph{heterogeneous} networks has only recently been explored \cite{fang2020effective}, and only for one query node. In addition, in our setting we aim to retrieve high-relevance groups, with ranked topics, authors and affiliations. 
We propose to retrieve relevant results for a user's query with two simple approaches. In the first, we compute the overlap between query facets $q$ and the top-$K$ salient facets $f$ for each group, and rank groups by normalized overlap size $\frac{|\{q: q \in f \}|}{|f|}$. In the second approach, we compute weighted PageRank scores \cite{xing2004weighted} over a graph with meta-nodes representing groups, and meta-edges as described earlier in this section. We do so separately for both types of edges: one for topical affinity, and the other for social proximity. At query time, we compute the average of these scores and the facet overlap score.

\section{Informal user studies and findings}
\label{sec:user}
We conduct preliminary user studies with four researchers and one practitioner. \textbf{P1} is a research scientist in virology, whose work also studies the Zika virus; \textbf{P2} is a postdoctoral fellow in the area of virology, working on viral infections and human antibody responses, and \textbf{P3} is a postdoctoral fellow and MD working primarily in Oncology. \textbf{P4} is a medical professional and PharmD. \textbf{P5} is a researcher working on viral diseases and proteins. %All participants used the system actively, searching for terms, topics and groups while following a think-aloud protocol in an hour-long session. We briefly summarize and discuss the main observations and themes that arose in our formative studies. For more details, see \ref{hope2020scisight}.

\textbf{Discovering unknown associations.} \ 
Based on interest in a the CR3022 antibody , \textbf{P2} searched for it with SciSight's collocation feature, finding ``very relevant associations'' and also ``two potentially surprising and interesting publications. I'm going to look into those papers.'' \textbf{P1} searched for cells linked to a type of cytopathic effect and found Calu-3 (a human lung cancer cell line), which led to ``spotting an interferon with relevant and interesting studies, very useful.'' \textbf{P4} discovered a link between broad-spectrum antivirals and MERS, providing a ``new and strongly relevant idea''. \textbf{P5} was able to find associations considered new and interesting between the TNF inflammatory cytokine and ERK1/2, a type of protein kinase (relevant to \textbf{P5}'s interest in cytokine profiles).

\textbf{Finding new groups and directions.} %\ Participants raised various ways in which they would use such a feature: Finding unknown labs or groups working on similar topics as potential competitors or collaborators, exploring around known groups to find related groups and directions, understanding what various groups are working on and how relevant they are. 
\textbf{P1} gave an example of a group (lab) using assays to identify which proteins antibodies bind to in order to to neutralize HIV, and connecting to other groups working on serum utilization for SARS to potentially collaborate. 
When searching for groups tied with a prominent scientist,
\textbf{P2} found a group associated with the scientist with recent work \cite{yuan2020highly} revealing a new direction regarding SARS-CoV-2 epitopes. \textbf{P2} also found a group in China with shared focus on epitopes but no social ties, revealing ``new perspectives that I would not have found otherwise'' on virus evolution.

\textbf{Limitations: more information and features.} \textbf{P1, P4} suggested that user-inputted concepts be combined with existing concepts/terms on-the-fly, and that edges could be removed by the user. \textbf{P3} and \textbf{P4} suggested ranking associations by ``measures of novelty'' to allow users to focus on emergent knowledge. Participants mentioned a diverse range of other entities to explore, e.g., patient weight, metabolic speed, drug dosages, vaccines, mutation mechanisms, and various techniques.

%  \textbf{Exploratory tools to complement search} \ All users mentioned the need for intuitive tools that can support exploratory needs unanswered by most search engines. \textbf{P4} frequently ``needs to look into research'', but this is difficult with standard search engines ``when you don't know what you don't know.''  Users also highlighted the need for a \textbf{user-friendly interface}, mentioning that SciSight's interface is ``easier to work with'' than other common tools (\textbf{P1}).

%\section{Discussion: Bibliometric visualization and search}
%\input{Discussion}
\section{Conclusion}
In this paper we presented SciSight, an evolving system for scientific literature search and exploration. We demonstrate SciSight's use on a large corpus of papers related to the COVID-19 pandemic and previous coronaviruses.  We use state-of-the-art scientific language models to extract entities such as proteins, drugs and diseases, and an overlapping community detection approach for identifying groups of researchers. To visualize groups we display group ``cards'' with a novel link scheme capturing topical and social affinities between communities, designed to identify socially disjoint groups working on similar topics. Preliminary user interviews suggest that SciSight can help complement standard search and may pave new research directions. In future work, we plan to conduct extensive studies to validate SciSight and better understand its potential and limitations.

%%
%% The next two lines define the bibliography style to be used, and
%% the bibliography file.
\bibliographystyle{acl_natbib}
\bibliography{sample-base}

%%
%% If your work has an appendix, this is the place to put it.
\end{document}